# The Contours of Crowd Capability

John Prpić
Beedie School of Business, Simon Fraser University
prpic@sfu.ca

Prashant Shukla
Beedie School of Business, Simon Fraser University
pshukla@sfu.ca

**Abstract**

In this work we use the theory of Crowd Capital as a lens to compare and contrast a number of IS tools currently in use by organizations for crowd-engagement purposes. In doing so, we contribute to both the practitioner and research domains. For the practitioner community, we provide decision-makers with a convenient and useful resource, in table-form, outlining in detail some of the differing potentialities stemming from crowd-engaging IS. For the research community, we begin to unpack some of the key properties of crowd-engaging IS, while simultaneously illustrating the usefulness of the Crowd Capability construct as means to understand the dynamics of crowd-engaging IS.

## 1. Introduction

The existence of dispersed knowledge has been a subject of inquiry for more than six decades [12]. Despite the longevity of this rich research tradition, the "knowledge problem" has remained largely unresolved both in research and practice, and remains "the central theoretical problem of all social science" [12]. However, in the 21$^{st}$ century, organizations are presented with opportunities through technology, to potentially benefit from the dispersed knowledge problem to some extent. One such opportunity is represented by the recent emergence of a variety of crowd-engaging information systems (IS).

In this vein, Crowdsourcing [5, 6] is being widely studied in numerous contexts, and the knowledge generated from these IS phenomena are well-documented [2, 13, 30]. At the same time, other organizations are leveraging dispersed knowledge by putting in place IS-applications such as Predication Markets [11] to gather large sample-size forecasts from within and without the organization. Similarly, we are also observing many organizations using IS-tools such as "Wikis" [14] to access the knowledge of dispersed populations within the boundaries of the organization. Further still, other organizations are applying gamification techniques [10, 24] to accumulate Citizen Science [8] knowledge from the public at large, through IS.

Among these seemingly disparate phenomena, a complex ecology of crowd-engaging IS has emerged, involving millions of people all around the world generating knowledge for organizations through IS. However, despite the obvious scale and reach of this emerging crowd-engagement paradigm, there are no examples of research (as far as we know) that systematically compares and contrasts a large variety of these existing crowd-engaging IS-tools in one work. Understanding this current state of affairs, we seek to address this significant research void by comparing and contrasting a number of the crowd-engaging forms of IS currently available to organizations.

To achieve this goal, we employ the Theory of Crowd Capital [22] as a lens to systematically structure our investigation of crowd-engaging IS. Employing this parsimonious lens, we first explain how Crowd Capital is generated through Crowd Capability in organizations. Taking this conceptual platform as a point of departure, in Section 3, we offer an array of examples of IS currently in use in modern practice to generate Crowd Capital. We compare and contrast these emerging IS techniques using the Crowd Capability construct to highlight some important choices that organizations face when entering the crowd-engagement fray. Decision-makers and researchers alike, to differentiate among the many extant methods of Crowd Capital generation, can use this comparison, which we term "The Contours of Crowd Capability". At the same time, our comparison also illustrates some important differences to be found in the internal organizational processes that accompany each form of crowd-engaging IS. In section 4, we conclude with a discussion of the limitations of our work.

## 2. Theoretical Background

From a resource based view (RBV) [3] and the knowledge-based view of the organization [26, 27], unique knowledge is viewed as valuable commodity for organizations, potentially giving them a competitive advantage over their competitors. Further, in recent times, innovation scholars have reasoned that organizations should give equal importance to internal and external knowledge for their R&D activities [7], and simultaneously, others have argued that the utilization of external knowledge gives organizations a competitive edge through decreased R&D costs [23]. Using these perspectives, Prpić and Shukla [22] bound and explain the dynamics and mechanisms that enable organizations to engage crowds through IS, and in doing so, supply a coherent and parsimonious model explaining how and why organizations engage in these disparate knowledge sources. The result is the Theory of Crowd Capital (see Figure #1 below adapted from [22]).

**Figure #1- The Theory of Crowd Capital**

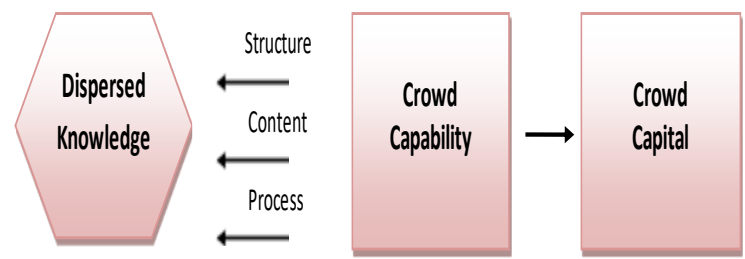

**Figure #1 – The dispersed knowledge of individuals is engaged and processed by the Crowd Capability of an organization, generating a heterogeneous Crowd Capital resource.**

The Theory of Crowd Capital suggests that a new form of heterogeneous knowledge resource is available to organizations that use IS to engage a crowd. The authors conceptualize that Crowd Capital is an organizational-level knowledge resource derived from the antecedent condition of dispersed knowledge [12], and further that an organization's Crowd Capability generates such knowledge. In their view, Crowd Capability is an organizational level capability, defined by the structure, content, and process of an organizations engagement with the dispersed knowledge of individuals—a Crowd [22]. The structure component of Crowd Capability is always an IS-mediated phenomenon and denotes the technological means employed to engage a Crowd population for the organization. The content dimension of Crowd Capability constitutes the knowledge, information or data that an organization seeks from a crowd population. Whereas, the process dimension of Crowd Capability defines the internal procedures that the organization will use to organize, filter, and integrate the incoming knowledge, information, and/or data.

Further, Prpić & Shukla [22] also delineate that the structure dimension of the Crowd Capability construct can be found to function in episodic or continuous forms, depending on the design of the IS used to engage dispersed knowledge. For example, Google's ReCaptcha, the Iowa Electronic Prediction market or Foldit; illustrate the episodic nature of Crowd Capability, where no community, collaboration, or interaction among the participants is needed through the IS, for Crowd Capital to be generated.

On the other hand, peer production [4] co-creation [20] and innovation communities [28] underscore the importance of social capital in efforts to engage an IS-mediated crowd. These efforts are "continuing" in nature, as there is interaction, community and collaboration between and among the participants using the IS to generate knowledge for the organization.

In the ensuing section of this work, we will use this theoretical platform to compare and contrast more than a dozen different IS tools currently in use for crowd-engagement.

## 3. The Contours of Crowd Capability

In this section, we present numerous examples of Crowd Capability currently in use to generate Crowd Capital for organizations. We discuss the nature of the crowd that these forms of crowd capability engage, and we further compare and contrast these IS application along the structure, content, and process dimensions of the Crowd Capability construct. Table #1 (see next page) summarizes the different "Contours of Crowd Capability" that we are observing in today's business environment. We will discuss each of the differentiating dimensions, in turn, below.

### 3.1 The Nature of a Crowd

Dispersed knowledge [12] is the antecedent condition of Crowd Capital generation for organizations, and in effect, dispersed knowledge is the state of nature in society, where all individuals have some unique knowledge relative to others. However, as is evident from the Table #1, the different IS tools analyzed here are designed to engage demonstrably different populations of participants.

Some efforts, like those of ReCaptcha and Wikipedia, engage the public at large, where contributions are, and can be made, by anyone. Other forms of IS, such as Crowdflower, M-Turk, and Hiretheworld also engage public crowds, though these applications do so in a "curated" manner. Curation [25] occurs when the individuals participating are "vetted" in one way or another as they participate through time. This "vetting" usually occurs through historical performance measures (such as leaderboards) or through techniques such as peer-evaluation, the award of badges for services rendered, or by the mutual assessment of participants [25]. Curation can thus serve as a barrier to entry for new participants to the community, while concomitantly providing signals [16, 25] that indicate the quality of the individual to those looking to engage a crowd. Although curation techniques have also been used specifically for content purposes in other settings [18, 17], for the purposes of our investigation we focus only on the curation of participants.

**Table 1- The Contours of Crowd Capability**

| Example of Crowd Capability in Use[1] | Nature of the Crowd Engaged | Crowd Capability - Structure | Crowd Capability - Content | Crowd Capability - Process |
|---|---|---|---|---|
| Google's ReCAPTCHA | Public Crowd – At Large | **Structure**: Web application structured in Episodic form. | **Content:** Images of text from analog books and newspapers. | **Process:** The web application aggregates the text inputted by Individuals into fully digitized works. |
| Wikipedia | Public Crowd – At Large | **Structure**: Web-based platform structured in Episodic form for contributors, and Continuing form for Editors. | **Content:** Encyclopedia entries. | **Process:** Edits contributed by individuals are monitored and judges by a community of editors. |
| Innocentive M-Turk Crowdflower Hiretheworld Kaggle 99Designs MobileWorks | Public Crowd – Curated | **Structure**: Web-based platforms structured in Continuing form. | **Content:** Variety of content categories available to be accessed by organizations, which are customizable to idiosyncratic organizational needs, including; Problem Solving, R&D, Microtasks, | **Process:** Organizations using the services provided by these intermediaries must internally process the knowledge that they receive through their own means. |
| NapkinLabs Imaginatik DataStation Lumenogic | Private Crowd – At Large | **Structure**: Software/Web applications structured in Continuing form. | **Content:** Content is customizable to meet the organizations idiosyncratic needs. | **Process:** Provide a variety of tools/features to assist the internal processing of incoming knowledge, including "Dashboards" for analysis. |
| BestBuy's TagTrade & Blue Shirt Nation | Private Crowd -- Curated | **Structure**: Web platforms structured in Continuing form for Blues Shirt Nation, and Episodic form for TagTrade. | **Content:** Market Research (TagTrade) & Internal Operations (Blue Shirt Nation) | **Process:** In TagTrade, knowledge is filtered through employee participation with simulated market mechanisms. With Blue Shirt Nation, employees, acting as a repository for future access, codify knowledge. |

---

[1] See Appendix #1 for a corresponding list of URL's

Furthermore, some forms of crowd-engaging IS are designed to engage private crowds, such as in the case of Best Buy's TagTrade & Blue Shirt Nation. In these examples, Best Buy leverages its own internal workforce as a private crowd of dispersed knowledge. Such crowds are private in the sense that Best Buy has sole ownership of the access to this collection of individuals. Further, such a private crowd is also curated in the sense that the employees are vetted by the organization before they are hired, providing both a barrier to entry for new participants and a signal of individual quality. Other forms of IS, like those found at NapkinLabs, DataStation, and Imaginatik, are designed to engage an organizations pre-existing crowd, for example through an organizations social media presence on Facebook or Twitter. Due to the fact that each organization controls the access to and ownership of its social media communities, we likewise consider that these crowds are private too, though with the distinction that it is a private-at large crowd, given that individual participation is not curated, and thus the general public can likely join such a crowd at any time.

Overall, when making a decision of which type of crowd-engaging IS to buy, rent or develop, an organization needs to consider and assess the nature of the crowd that it would like to engage, realizing that not all forms of crowd-engaging IS are created equal in this regard.

## 3.2 The Crowd Capability Dimensions: Structure

As we have learned thus far in this exposition, Crowd Capability includes three dimensions that need to be considered before an organization can engage dispersed knowledge. The structure dimension details the form of IT that the organization will use to engage a crowd, the content dimension details the specific type of knowledge that the organization seeks from a crowd, and the process dimension outlines the internal processes that the organization will employ to filter, organize, and purpose the knowledge that is received from a crowd. In this regard, and as is evident from Table #1, there is a great deal of variety to be found across all three dimensions of Crowd Capability in practice today.

The structure dimension of Crowd Capability is where an organization uses IS to engage a crowd. As we can see from Table #1, some organizations are using mobile software applications like MobileWorks [19], others are using software with the web (e.g. ReCaptcha), others still have created web-platforms (Wikipedia), while others (like M-Turk) use web platforms to offer services of crowd intermediation [1]. Beyond these extant examples that we focus upon, there are other forms of IS, including, bots, sensor technologies, 3D printers etc. that could also potentially be productively employed (either independently or in combination) to engage a crowd, though for the purposes of this paper, we limit our analysis to the forms detailed above. Nonetheless, this current variety found in the structure dimension of Crowd Capability indicates that organizations have many IT options available to engage dispersed knowledge through IS.

## 3.3 The Crowd Capability Dimensions: Content

The content dimension of Crowd Capability also displays a wide variety of different knowledge needs/goals in our comparison in Table #1. We can see that some forms of IS are used to generate knowledge in a literal sense, as in the case of Wikipedia and its encyclopedia entries. Others are targeted at generating ideas and creativity, as in the case of Hiretheworld and 99Designs. Further still, others like Kaggle are solving specific problems for organizations; while endeavours like Innocentive are generating R&D. Intermediation platforms like M-Turk and Crowdflower provide ready and willing labour for organizations that can perform a variety of tasks (perhaps any of those already mentioned), though thus far the individuals at such intermediation services are thought to excel at microtasks such as the translation of documents, labelling photos, and participating in surveys [19].

From our perspective, Google's ReCaptcha is a particularly interesting application of the Crowd Capability content dimension. Given that Google seeks to digitize books through the automation of microtasks, and said microtasks simultaneously serve a dual purpose -- to reduce spam and digitize books at the same time--, it appears that with ReCaptcha, Google serves to combine two Crowd Capability content tasks (spam reduction and text digitization) into one Crowd Capability IS structure, which is the first of its kind to achieve such a feat from our perspective.

Overall, we hope that you can see that organizations have a lot of existing options to choose from when considering the types of content that they desire to access from crowds.

## 3.4 The Crowd Capability Dimensions: Process

In terms of the process dimension of Crowd Capability, here too we find a variety of approaches in Table #1. Because this dimension delineates the internal processes that an organization will institute or employ, to filter, organize, and purpose the knowledge gained from a crowd, the process dimension can in essence be thought of as the "last mile" of Crowd Capital creation.

As we see from Table #1, some existing forms of Crowd Capability like M-Turk and Crowdflower, provide organizations with little or no support for the internal processing of incoming crowd knowledge. On the other hand, other forms of Crowd Capability such as ReCaptcha involve some significant pre-processing work, in the sense that the bits of text that are undecipherable by OCR have to be transformed into digital images before they can be used in the ReCaptcha system.

Some Crowd Capability applications, such as NapkinLabs for example, in effect specialize in the process aspect of Crowd Capability. In this case, they offer customized solutions for the internal processing of incoming knowledge, through features of the application such as "Insight Dashboards" that are specifically designed to aid your organization's analysis and use of incoming knowledge.

Furthermore, Crowd Capability applications can use well-known incentives such as pricing mechanisms, (as is common with Prediction Markets), to persuade individuals to process the knowledge generated from a crowd [11]. In this vein, the example of Best Buy and its TagTrade process is a powerful exemplar.

In our view, of particular interest along the process dimension of Crowd Capability is the example that Wikipedia provides. In the Wikipedia system relatively small crowds of specialists evaluate and approve the knowledge generated by the larger public crowd. This fact is impressive, given that both sets of Wikipedia crowds are volunteers, but perhaps more importantly, it signals the opportunity that organizations may benefit from using one crowd to process the knowledge generated from other different crowds. From our perspective, it would be really interesting to see if an organization could implement a Best Buy type of private crowd to process the knowledge from a large public crowd, derived for example from M-Turk or Crowdflower. Alternatively, we feel the reverse may also be interesting, in that an organization could use an M-Turk type public crowd to process and evaluate the knowledge from a Best Buy type of private crowd. Whatever the case may be, in our view, it may be that the future of Crowd Capability lies in part in using multiple crowds in parallel.

And finally in terms of the process dimension of Crowd Capability, it is important to note that other forms of extant IS (which were not created for crowd-engagement) such as data mining [9] and business intelligence [29] applications, might also be fruitfully employed by the organization for the purposes of processing incoming crowd knowledge. Though we are as yet unaware of any extant situations employing such a configuration in the domain of crowd engagement, it may well be that such "layering" of applications has significant bearing on the economics of processing incoming crowd knowledge.

## 3.5 Episodic Vs. Continuing Structure

The final characteristic, upon which we differentiate the different forms of Crowd Capability in Table #1, is a subset of the structure dimension. Here, we draw upon the important distinction made by Prpić & Shukla [22], of Episodic vs. Continuing Crowd Capability structure. As mentioned earlier in this work, an Episodic structure is a form of IS that does not need collaboration, cooperation, interaction or relationships among the engaged participants for the knowledge resource to be generated. Google's ReCaptcha is a leading example of this type of structure, as the very many contributors (approximately 200 million ReCaptcha's per day are typed by users, equalling about 500,000 hours of work per day[2]) never interact with one another.

On the other hand, Continuing Crowd Capability structure uses collaboration and relationships among the participants, in one form or another, to generate knowledge resources. A good exemplar in this realm is Best Buy's Blue Shirt Nation, which relies on a volunteer community of internal employees to exchange knowledge with one another on various topics.

Once more, we find that Wikipedia is an interesting example along this dichotomy too, in that it implements both forms of Crowd Capability structure simultaneously. Wikipedia implements an Episodic structure by allowing anybody to contribute to the encyclopedia in a "one-off" manner, yet, at the same time, Wikipedia employs a Continuing structure too, through the community of editors that monitor and approve these episodic contributions.

---
[2] https://plus.google.com/+PeterHDiamandis/posts/K9TWkL3miDD

In terms of the knowledge economics of this structure dichotomy, we feel that the different combinations presented here would similarly entail very different economic repercussions. For support in this regard, we point to the simple fact that some episodic structures like ReCaptcha "automate" the processing of incoming crowd knowledge, whereas most, if not all, continuing structures require human processing of the incoming knowledge. We would thus expect that such structural differences, encoded in the IS, would have a major bearing on both the timing and amount of cash flows associated with each structure.

Overall, it is very important for organizations to understand the episodic-continuing didactic of Crowd Capability structure, as most practitioners and researchers falsely assume that community, collaboration, and interactions within a crowd is necessary to generate knowledge therein. As we have illustrated, continuing Crowd Capability structure is but one of the possible options available to organizations considering to enter the crowd fray.

**4: Limitations & Conclusion**

In this work we use the theory of Crowd Capital as a lens to compare and contrast more than a dozen emerging IS tools currently in use by organizations for crowd-engagement purposes. We systematically employ the dimensions of the Crowd Capability construct to differentiate among these emerging forms of IS, and thus begin to outline the contours of Crowd Capability currently in use for crowd engagement by organizations. In terms of IS structure, we find that organizations are using mobile software applications, web-based software applications, web platforms, and web platforms offered as intermediation services. We further find that these IS tools can be considered to be in either an episodic or continuing form.

In parallel, we also begin to explore the nature of the different crowds that organizations can engage through the aforementioned IS tools. We find that some forms of the IS analyzed here targets private crowds, while others engage public crowds. Further, we find that in both cases, said crowds (either private or public) can be found to occur in either "curated" or "at large" forms, where in curated form individual crowd participants are vetted to some degree.

Like all research, our work here has limitations. Our investigation is very far from exhaustive, in the sense that we only consider fourteen different forms of IS tools currently used for crowd engagement. Most likely, there are forms of crowd engaging IS of which we are currently unaware, and further, within our categorization there are certainly many more examples that could have been included. Further, our analysis method is ex post, and although we urge you to investigate the forms of IS that we point to (see Appendix #1) we make absolutely no claims to the correctness of our analysis. Rather it is our hope that our work here is a decent starting point in this vein.

Further still, due to the nascent and continuously emerging nature of our subject matter, there is not a rich and deep tradition of research for us to ground our claims. Although wherever possible we employ the extant literature faithfully, the lack of extant research signals the need for caution in accepting our results.

Despite these limitations, we believe that our work is indeed a useful starting point. We systematically use the Theory of Crowd Capital to bound and limit our investigation, and in doing so, draw-out some useful dichotomies in the crowd engaging IS domain. In doing so, we raise interesting questions for future research, and we look forward to future research that investigates the intersection between private and public crowds, or which investigates the relative efficacy of episodic structures versus continuing forms. Further, we begin to address the nature of the crowd that different forms of IS engage, and we believe that this will prove to be a rich vein of research investigating the traits and relative merits of curated crowds vs. at large crowds etc.

Similarly, we feel that our work is a very useful resource for the practitioner community, especially for those organizations who are considering beginning some crowd engagement endeavors. Our work supplies decision makers in these organizations with a systematic starting point, which highlights some key decision issues that should be considered, both strategically and operationally, before beginning to implement Crowd Capability, and thus generating Crowd Capital.

## 6. Appendix #1

| | |
|---|---|
| 99Designs - | http://99designs.ca/ |
| Best Buy | http://online.wsj.com/article/SB122152452811139909.html |
| Crowdflower - | http://crowdflower.com/ |
| DataStation - | http://www.datastation.com/ |
| Hiretheworld - | https://www.hiretheworld.com/ |
| Imaginatik - | http://www.imaginatik.com/ |
| Innocentive - | http://www.innocentive.com |
| Kaggle - | http://www.kaggle.com/ |
| Lumenogic - | http://www.lumenogic.com/www/index.html |
| MobileWorks - | https://www.mobileworks.com/ |
| M-Turk - | https://www.mturk.com/mturk/welcome |
| NapkinLabs - | http://napkinlabs.com/ |
| ReCaptcha - | http://www.google.com/recaptcha |
| Wikipedia - | http://www.wikipedia.org/ |